\documentclass{aa}
\usepackage{graphics}

\begin{document}
	\title{Theoretical interpretation of the X-ray properties of GRB960720}
	\author{F. Daigne \and R. Mochkovitch}
	\institute{Institut d'Astrophysique de Paris, 
	CNRS, 98 bis boulevard Arago, 75014 Paris France}
	\date{Received **.**.**; accepted **.**.**}
	\maketitle
	
	\begin{abstract}
	BeppoSAX observations of the single pulse burst GRB960720 have allowed 
	a detailed study of its X-ray properties: pulse width in
	different energy bands, spectral evolution from 2 to 700 keV, etc. We show 
	that the early (0--5s) X-ray emission and the gamma-rays are well
	explained by 
	internal shocks in a relativistic wind while the late (5--20s) X-ray emission
	could come from the reverse shock generated in the wind when it interacts
	with the external medium. The results for a medium of uniform density 
	are compared to the observations.
	\end{abstract}
	\section{Introduction}
The gamma-ray emission from GRBs is probably
produced by internal shocks in a relativistic wind whereas the afterglow (from X-rays to radio bands) is due to the external shock, i.e. the forward shock propagating in the ISM because of its interaction with the wind (Rees and M\'esz\'aros, \cite{rees} ; Wijers et al., \cite{wijers}). Simultaneously, a reverse shock propagates in the wind itself. We illustrate here the possible contribution of this reverse shock to an X-ray emission perduring immediately after the gamma-rays. Such an emission has for been observed in the first GRB detected by Beppo-SAX, GRB960720. We use the detailed X-ray data made available by SAX for this burst (Piro et al., \cite{piro}) to make a comparison with our theoretical results.

	\section{The internal shocks}
GRB960720 has been observed both by BATSE and Beppo-SAX. It is a single-pulse burst, with a ``FRED'' profile. Its duration in the 50-700 keV band is around 2--3 s
but the X-ray emission lasts longer: Piro et al. (\cite{piro}) show that the power-law between the width of the pulse and the energy (already known in the gamma-ray range) is observed down to 2 keV. They find $W(E) \propto E^{-0.46}$.\\
We use a simple model to simulate internal shocks and build synthetic bursts: all pressure waves are neglected so that we consider only direct collisions between solid layers. In the shocked material, the magnetic field reaches equipartition values (10 -- 1000 G) and the Lorentz factor of the electrons is obtained from the dissipated energy per proton $\epsilon$ using the formula given by Bykov and M\'esz\'aros (\cite{bykov}) who suppose that only a fraction $\zeta$ of the electrons is accelerated:
		\begin{equation}
\Gamma_{e} \sim \left[ \frac{\alpha_{M}}{\zeta} \frac{\epsilon}{m_{e} c^{2}}\right]^{\alpha}\, {\rm (} \alpha_{M} \sim 0.1 \to 1\ {\rm and}\ 1 \le \alpha \le 1.5 {\rm)}
		\end{equation}
For $\zeta \sim 1$ the usual equipartition assumption yields values of $\Gamma_{e}$ of a few hundreds: the gamma-ray emission is due to inverse Compton scattering on the synchrotron photons. Smaller values for the fraction of accelerated electrons ($\zeta < 10^{-2}$) lead to larger Lorenz factors ($\Gamma_{e}$ of a few thousands) so that the gamma-rays are directly produced by the synchrotron process, which is the assumption made here. Internal shocks have been shown to successfully reproduce the main temporal and spectral properties of GRBs (Daigne and Mochkovitch, \cite{daigne}).\\
	\begin{figure}[!b]
\begin{center}
\resizebox{!}{4.5cm}{\includegraphics{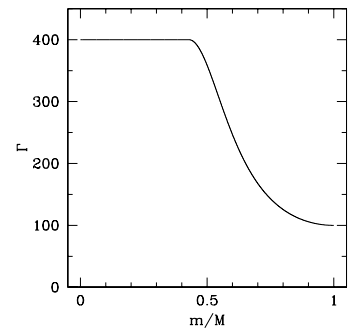}}
\end{center}
\vspace{-0.5cm}
\caption{Initial distribution of the Lorentz factor in the wind.}
\label{FigGamma}
	\end{figure}
We model GRB960720 with a wind emitted during 4 s and consisting in a slow and a rapid part of equal mass (see figure \ref{FigGamma}).
Two internal shocks are generated and we sum both contributions to the emission to construct the synthetic burst.  The profile in the SAX 50-700 keV band looks very similar to GRB960720 as can be seen in figure \ref{FigISM}. However the X-ray emission does not last long enough so that the power-law relating $W(E)$ and $E$ is not reproduced in this spectral range.

	\section{Effect of a medium of uniform density}
\begin{figure*}[t]
\resizebox{!}{6cm}{\includegraphics{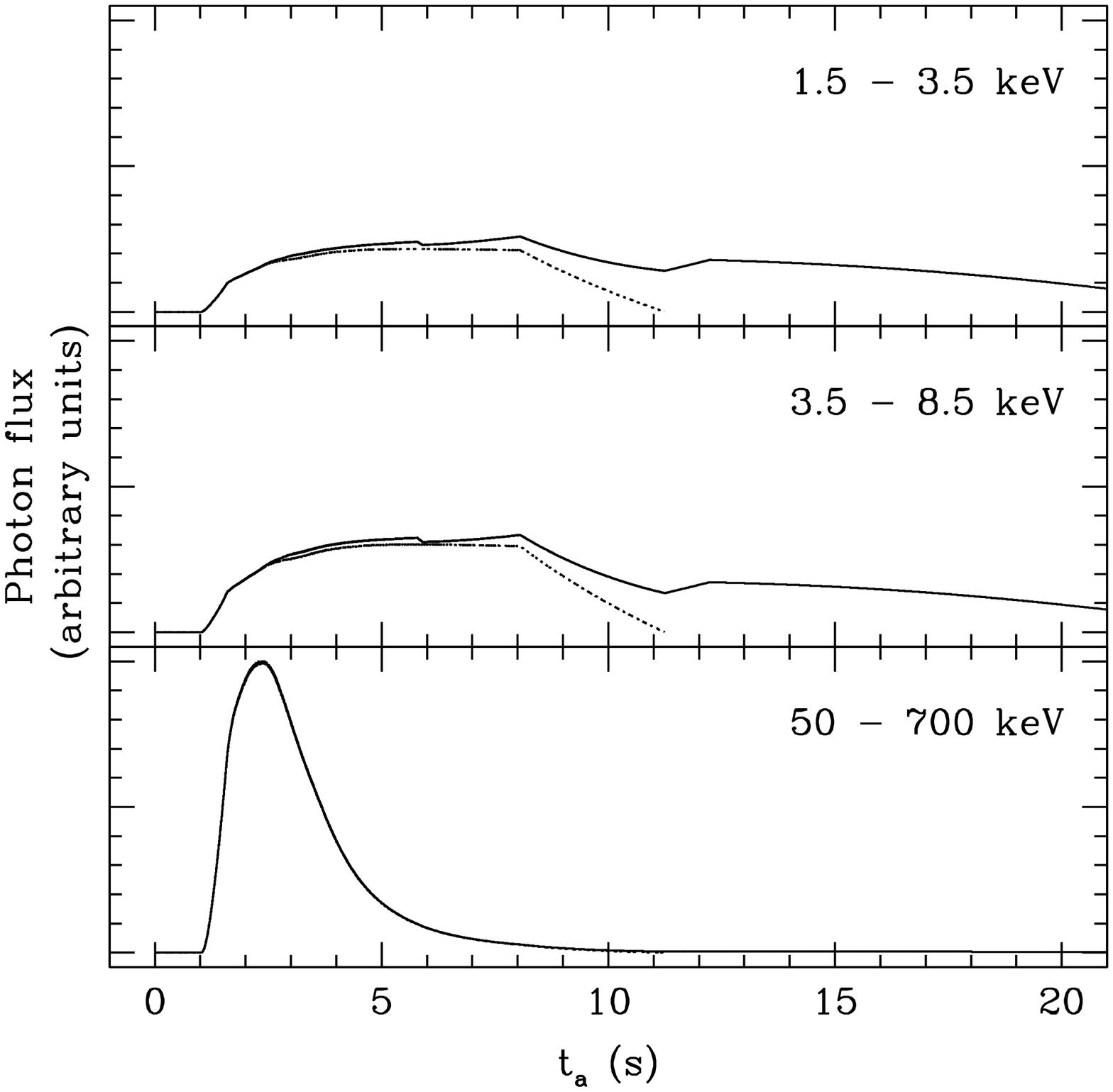}}
\resizebox{!}{6cm}{\includegraphics{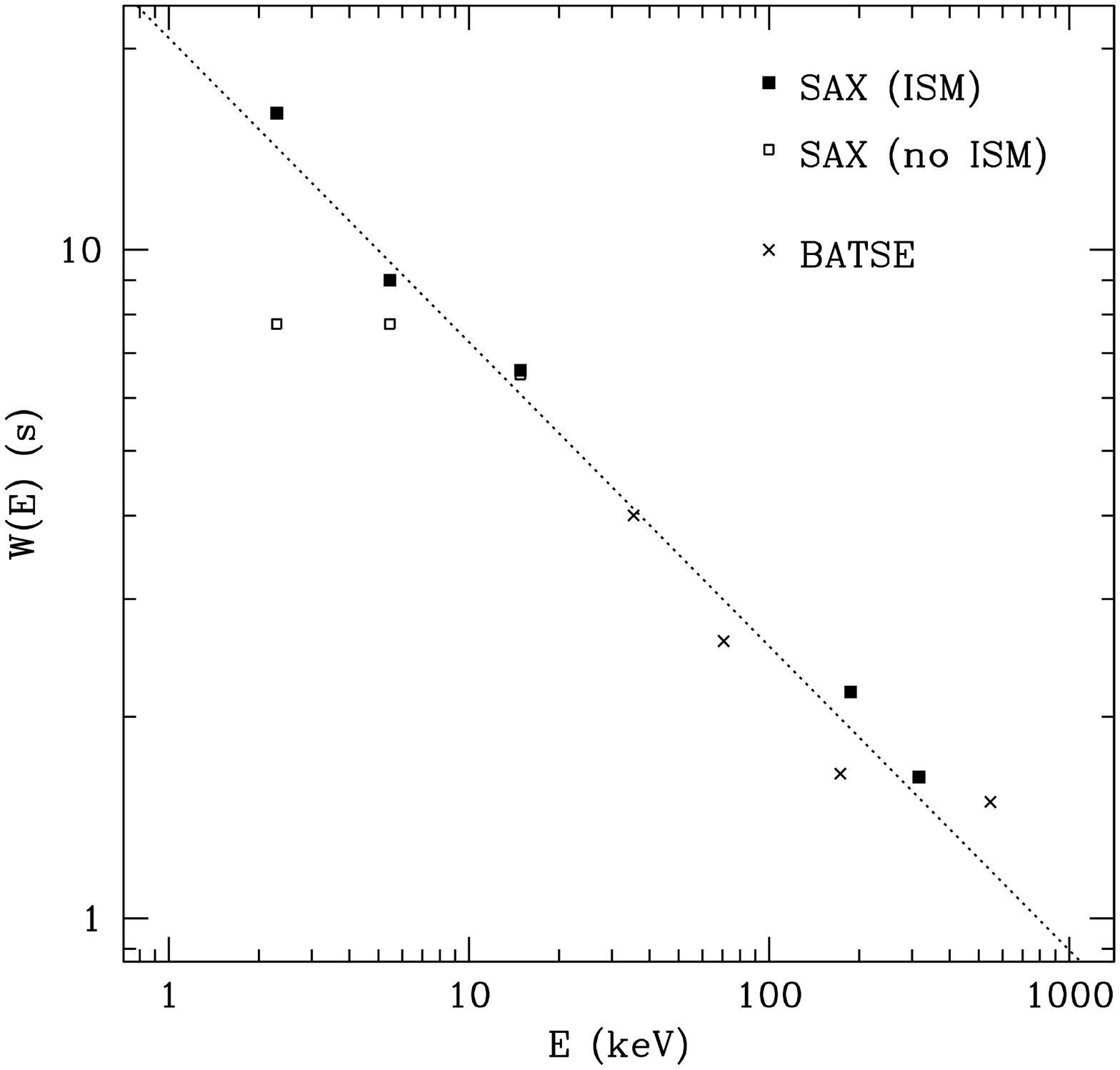}}
\resizebox{!}{6cm}{\includegraphics{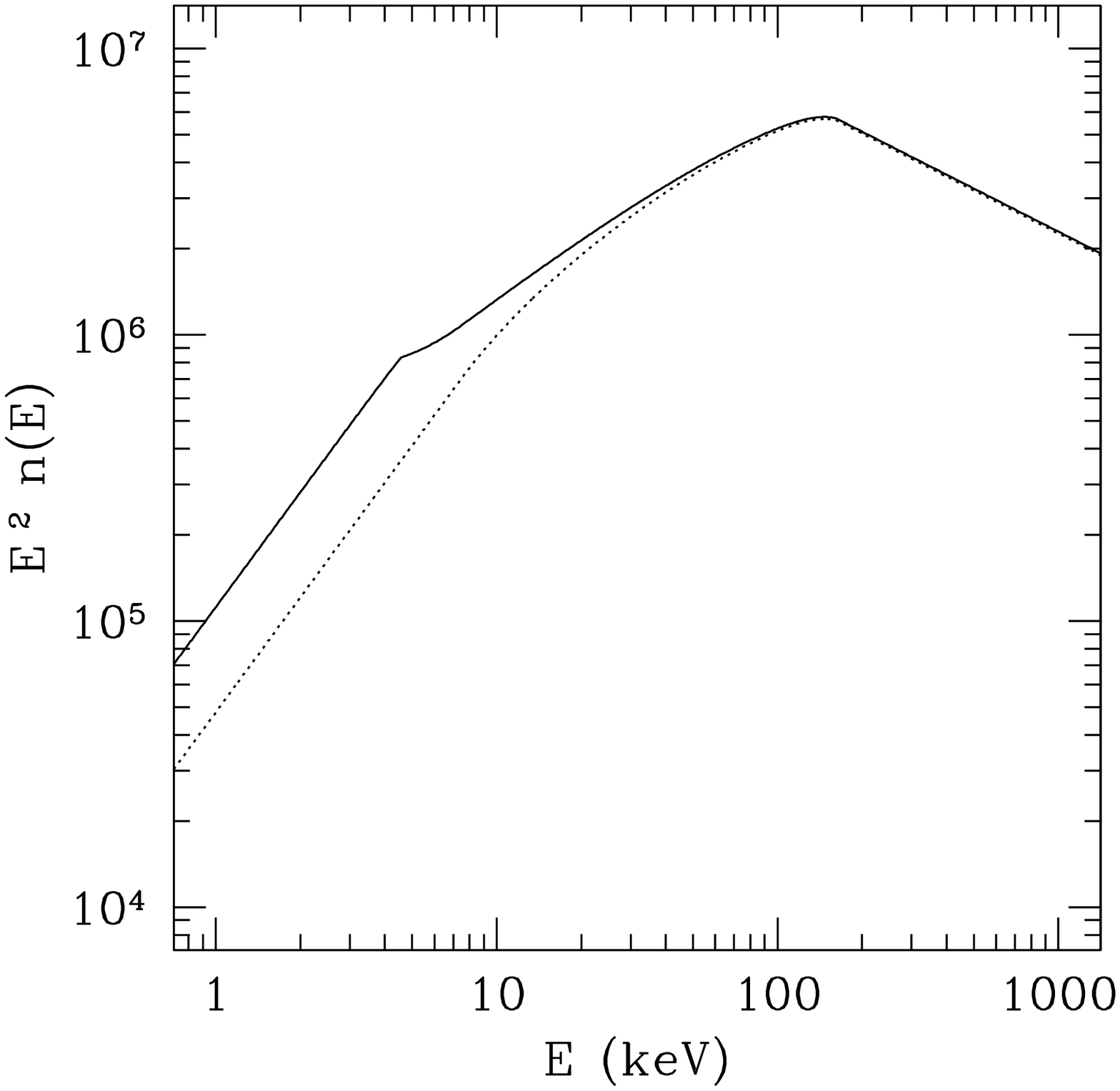}}
\vspace{-0.6cm}
\caption{Results obtained for a ratio $E^{inj}_{52} / n = 0.5$. For $E_{\gamma}=10^{51}/4\pi\ erg/sr$ and $f_{\gamma} = 0.05$ it corresponds to $n=4\ cm^{-3}$. \textit{Left panel :} Profiles of the synthetic burst in the X-- and gamma--ray bands. The full line takes into account the presence of the ISM whereas the dotted line only shows the contribution of the internal shocks.
\textit{Middle panel :} Pulse width as a function of energy. The dotted line corresponds to $W(E) \propto E^{-0.45}$.
\textit{Right pannel :} Burst spectrum. The full line takes into account the ISM and the dashed line does not.}
\label{FigISM}
\end{figure*}

We now consider the effect of the ISM, whose density $n$ is supposed to be uniform. The external shock (forward shock) produces the afterglow. 
Simultaneously a reverse shock propagates into the wind.
Its strength is comparable to that of the internal shocks, which are midly relativistic while the external shock is initially very strong and relativistic. We therefore adopt the same assumptions to compute the emission of the shocked material behind the reverse and the internal shocks.\\
It is possible to derive a critical ratio of the total energy injected over the density (see Sari and Piran, \cite{sari}) :
\begin{equation}
\left.\frac{E^{inj}}{n}\right|_{crit} \sim \frac{4 \pi}{3} m_{p} c^{5} \bar{\Gamma}^{8} T^{3}
\end{equation}
for which the reverse shock will interfere with the internal shocks. Injecting the typical values used in our example (Lorentz factor $\bar{\Gamma} \sim 300$ and duration $T = 4 s$) gives
\begin{equation}
\left.\frac{E^{inj}_{52}}{n}\right|_{crit} \sim 0.071 \bar{\Gamma}_{300}^{8} T_{4}^{3}\, ,
\end{equation}
where $E^{inj}_{52} = \frac{E^{inj}}{10^{52}/ 4 \pi\ erg/sr}$. 
Assuming an efficiency $f_{\gamma}$ for the conversion of wind energy into gamma-rays by internal shocks allows to obtain $E^{inj}=\frac{E_{\gamma}}{f_{\gamma}}$ from the observed burst energy $E_{\gamma}$.\\
The total energy being fixed, if the density is smaller than the critical value, the reverse shock does not contribute
in gamma-rays and only produces a delayed X--ray emission with an intensity which increases with $n$.
If the density reaches or exceeds the critical value, the X--ray and the gamma-ray profiles become very affected.\\
We have represented in figure \ref{FigISM} the profiles for a ratio $E^{inj}_{52} / n = 0.5$ ($n \sim\ 0.15 n_{crit}$).
The gamma-ray profile is unchanged and the X-ray profiles are improved. The corresponding $E$ -- $W(E)$ diagram in figure \ref{FigISM} shows a well reproduced power-law over the complete energy range with an index $-0.45$ in agreement with the observations of GRB 960720. In the spectrum the contribution of the reverse shock to the late emission appears like a X--ray plateau, which is observed (and is even more extended) in GRB960720.
	\section{Conclusions}
The reverse shock propagating into the wind following its interaction with the ISM has been shown to produce, for a sufficiently dense medium, a late X-ray emission. In the case of GRB960720, whose gamma-ray properties are well explained by internal shocks, taking into account this additional contribution greatly improves the X-ray profiles (the power-law between the profile width and the energy is reproduced) and the spectrum.\\
It is now important to study the case of more complex environments. In hypernova models, the progenitor is a massive Wolf-Rayet star with a dense wind ($\dot{M} \sim 3\ 10^{-5}\ M_{\odot}/yr$ and $v_{\infty} \sim 2000\ km/s$ typically) leading to high densities ($n \sim \frac{\dot{M}}{4 \pi r^{2} m_{p} v_{\infty}}$) near the source. Our first calculations show that in this case even the gamma-ray profile could be strongly affected by the reverse shock, which may represent a potential problem.\\


\vspace{-0.4cm}
	\begin{thebibliography}{}
		\bibitem[1996]{bykov} Bykov A.  and M\'esz\'aros P., 1996, ApJ
		461, L37
		\bibitem[1998]{daigne} Daigne F. and Mochkovitch R., 1998, 
		MNRAS 296, 275
		\bibitem[1998]{piro} Piro L. et al., 1998, A\&A 329, 906
		\bibitem[1994]{rees} Rees M.J. and M\'esz\'aros P., 1994, ApJ 430, L93
		\bibitem[1995]{sari} Sari R. and Piran T., 1995, ApJ 455, L143
		\bibitem[1997]{wijers} Wijers R.A.M.J., Rees M.J. and M\'esz\'aros P., 1997, MNRAS 288, L51
	\end{thebibliography}
\end{document}